\documentclass[10pt,twoside]{article}

\usepackage[english] {babel}

\usepackage {graphicx}
\usepackage {graphics}
\usepackage {floatflt}
\usepackage {latexsym}
\usepackage {caption3} %Пакет для MikTex 2.6, где содержатся команда для замены двоеточия на точку в подписи к рис., например, Рис.1: -> Рис.1.
\usepackage[labelsep=period]{caption}[2007/11/04 v.3.1e]
\usepackage {amssymb}

\def\noi{\noindent}

\renewcommand{\thesubsubsection}%
        {\arabic{section}.\arabic{subsection}.\arabic{subsubsection}.}

\newcommand{\heads}[2]{\markboth{\protect\small\it #1}{\protect\small\it #2}}

\newcommand{\Arthead}[5]{ \setcounter{page}{#4}\thispagestyle{empty}\noi
    \unitlength=1pt \begin{picture}(500,40)

        \put(0,58){\shortstack[l]{\small\it Gravitation \& Cosmology,
                        \small\rm Vol. #1 (#2), #3, pp. #4--#5    \\%No. #3, pp. #4--#5    \\
        \footnotesize {Proceedings of the 5th Inernational Conference on Gravitation and Astrophysics of Asian-Pacific Countries (ICGA-2001), }    \\
\footnotesize {Moscow, 1-7 October 2001}   \\
\footnotesize\copyright \ #2 \ Russian Gravitational Society} }

    \end{picture}
	 }     		%%% #1: volume; #2: year; #3: issue; #4-#5: pages
\def\prepno#1#2
    {\thispagestyle{empty}
    \noi    \unitlength=1mm
    	\begin{picture}(178,10)
            \put(177,15){\llap{\bf #1}}
            \put(177,10){\llap{\small\rm #2}}
        \end{picture}
          }             %%% #1: prepr.no.; #2: journal-ref

\newcommand{\Title}[1]{\noi {\uppercase{\Large #1}}     }%\\}
\newcommand{\Author}[2]{\noi{\large\bf #1}\\[2ex]\noindent{\it #2}   }%\\}

\newcommand{\Abstract}[1]{\vskip 2mm \begin{center}
        \parbox{16.4cm}{\small\noi #1} \end{center}\medskip}

\newcommand{\foom}[1]{\protect\footnotemark[#1]}

\newcommand{\email}[2]{\footnotetext[#1]{e-mail: #2}
		\addtocounter{footnote}{1}}

\heads{A.M.Baranov}
      {Interior Static Stellar Model with Electric Charge as an Oscillator}

\begin{document}
\twocolumn 
[
\Arthead{8}{2002}{Supplement II}{10}{11}

\Title{INTERIOR STATIC STELLAR MODEL WITH ELECTRIC CHARGE %\\

\vspace{0.2cm}
AS AN OSCILLATOR \foom 1}%\\

\vspace{.5cm}
   \Author{A.M.Baranov\foom 2 }   
{\it Dep. of Theoretical Physics, Krasnoyarsk State University,
79 Svobodny Av., Krasnoyarsk, 660041, Russia}

%\vspace{.3cm}
%{\it Received 11 October 2005}
%
\vspace{.5cm}
\Abstract
{A model approach to the description of static stars filled with a charged Pascal perfect fluid within the framework of general relativity is investigated. The metric is written in Bondi's radiation coordinates. The gravitational equations are reduced to a nonlinear oscillator equation after transfomation to a new variable as a function of the radial coordinate. It is shown that in this case exact solutions of the Einstein-Maxwell equations for a concrete energy density distribution law of the charged fluid may be obtained as solution of the harmonic oscillator equation. }
\vspace{.5cm}
]

\footnotetext[1]{Talk presented at the 5th Int. Conf. on Grav. and Astrophys. of Acian-Pacific Countries (ICGA-5), Moscow, 2001.} 
\email 2 {alex\_m\_bar@mail.ru}

In general relativity, together with Schwarzschild's exterior solution (see, e.g., [1]), which describes the gravitational static field of spherical objects, there is a solution of the gravitational equations for an exterior field of a massive spherical body with an electrical charge, well known as the Reissner-Nordstr$\ddot{o}$m \linebreak
solution (see, e.g., [2]). However, as to a similar problem of finding of the solution for an interior field of an electrified body, there are quite clear difficultes of the solution of the Einstein-Maxwell nonlinear differential equations in a medium.

Let us consider an interior static stellar model having some distribution of the electric charge, within the framework of general relativity. The exterior gravitational field of such a star is the Reissner-Nordstr$\ddot{o}$m solution [2]. The metric is written in Bondi's radiation coordinates:

$$
\hspace {3mm}ds^2 = F(r)dt^2 +2L(r)dtdr-r^2(d{\theta}^2 + sin{\theta}^2d{\varphi}^2),
\eqno{(1)}
$$
where $t$ is a time coordinate, $r,$ is a radial coordinate, 
$\theta,$ and $\varphi$  are angular variables; the Greek indices take the values $\;0,1,2,3;$ $\; F(r)$ and $L(r)$ are metric functions to be found. They depend only on the radial variable. The velocity of light and the Newtonian gravitational constant are chosen to be equal to unity.

The Einstein equations for the interior fielf are written with a source for a charged substance as 

$$
\hspace {-30mm}G_{\alpha \beta} = R_{\alpha \beta}-\frac{1}{2}g_{\alpha \beta}R =
-\hbox{\ae} T_{\alpha \beta}, 
\eqno{(2)}
$$
where $G_{\alpha \beta}$ is the Einstein tensor, $R_{\alpha \beta}$ is the Ricci tensor, 
$R = R^{\alpha}_{\cdot\alpha}$ is the scalar curvature, $\varkappa =8\pi$ is Einstein's constant, 
$g_{\alpha \beta}$ is the metric tensor. The resulting energy-momentum tensor of the charged substance 
$T_{\alpha \beta}$ is taken here as a sum of the energymomentum tensor of the Pascal neutral perfect fluid 

$$
\hspace {-36mm}T^{fluid}_{\alpha \beta} = (\mu + p) u_{\alpha} u_{\beta} -
pg_{\alpha \beta}
\eqno{(3)}
$$
and the electromagnetic energy-momentum tensor

$$
\hspace {-20mm}T^{el-mag}_{\alpha \beta} =
\displaystyle\frac{1}{4\pi}(-F_{\alpha \mu}F_{\beta}^{\mu}
+g_{\alpha \beta}F_{\mu \nu}F^{\mu \nu}),
\eqno{(4)}
$$
where $\mu(r)$ is the density, $\;p$ is the pressure, $\;u^{\alpha} = dx^{\alpha}/ds $ is the 4-velocity, $\;F_{\alpha \beta}$ is the electromagnetic field tensor, all functions depend only on the radial variable. 

The Maxwell equations (the second pair) in our case become

$$
\hspace {-24mm}{F^{\mu \nu}}_{; \nu} =
\displaystyle\frac{1}{\sqrt{-g}}(\sqrt{-g} F^{\mu \nu})_{, \nu} =
-4\pi j^{\mu}, 
\eqno{(5)}
$$
where $g$ is the determinant of the metric tensor, $j^{\mu}$ is the electrical current density. 

The statics nature of the system allows one to introduce a fixed comoving frame of reference with 

$$
\hspace {2mm}u^{\mu}= \displaystyle\frac{\delta^{\mu}_0}{\sqrt{g_{00}}} =
\displaystyle\frac{\delta^{\mu}_0}{\sqrt{F(r)}};\;\;\;\; u_{\mu}=
\displaystyle\frac{g_{\mu_0}}{\sqrt{g_{00}}} =
\displaystyle\frac{g_{\mu_0}}{\sqrt{F(r)}}
\eqno{(6)}
$$
and the physically observable quantities: the mass-energy density, the electrical charge density, the electric intensity (radial component) and the electric field energy density will be written accordingly as

$$
\hspace {2mm}\mu_{phys} = \mu(r); \;\;\; \rho_{phys} \equiv \rho = j^{\mu} u_{\mu} =
j^0 \sqrt{g_{00}};
\eqno{(7.1)}
$$
$$
\hspace {2mm}E_{phys} \equiv E_1 =
\displaystyle\frac{F_{01}}{\sqrt{g_{00}}}=
\displaystyle\frac{E}{\sqrt{g_{00}}};\;
\; W_{el} = \displaystyle\frac{E^2}{8\pi {g_{01}}^2}. 
\eqno{(7.1)}
$$

As a result, the set of Einstein-Maxwell equations is reduced to the following set of nonlinear differential equations:

$$
\hspace {-20mm}\displaystyle\frac{F}{xL^2} (ln L)^{\prime} = \hbox{\ae} (p+\mu); 
\eqno{(8.1)}
$$
$$
\hspace {-16mm}\displaystyle\frac{F}{xL^2} (ln L)^{\prime} -
 \displaystyle\frac{1}{2L^2}\left[F^{\prime\prime} +
 \displaystyle\frac{2}{x} F^{\prime}
- {F^{\prime}}(ln L)^{\prime}\right] 
$$
$$
= -\varkappa (p + W_{el}); 
\eqno{(8.2)}
$$

$$
\hspace {-30mm}\displaystyle\frac{1}{x^2}(-1+ \displaystyle\frac{F}{L^2}+
\displaystyle\frac{xF^{\prime}}{L^2} -\displaystyle\frac{xF (ln
L)^{\prime}}{L^2} 
$$
$$
= -\varkappa \left[\displaystyle\frac{(\mu -p)}{2} + W_{el}\right]; 
\eqno{(8.3)}
$$
$$
\hspace {-42mm} \left(\displaystyle\frac{x^2 E}{L}\right)^{\prime} =
4\pi R \rho x^2 \displaystyle\frac{L}{\sqrt{F}} , 
\eqno{(8.4)}
$$
where the prime denotes a derivative with respect to $ x = r/R ,\;$ 
$0 \leq x \leq 1 $, $R$ is the radius of the star, $\;W_{el} = {E^2}/8 \pi L^2,\;$
$E$ is the electric field intensity. 

By the replacement $\varepsilon = F/L^2 $ it is possible to derive a linear second-order ordinary differential equation with variable coefficients from the first three Einstein equations by eliminating the presure and the mass density: 

$$
\hspace {-40mm}G^{\prime \prime}  + f (x) C^{\prime} + g (x) G = 0, 
\eqno{(9)}
$$
where $ f(x)=(ln \varphi)^{\prime}, \; \varphi = \sqrt {\varepsilon}/x, \;$
$g(x) = [2(1-\varepsilon) +x \varepsilon^{\prime}]/(2 x^2 \varepsilon)
-2\chi W_{el}/\varepsilon; \;$ $\chi = \varkappa R^2.$

For the new variable $\zeta = \zeta(x)$ introduced by the relation 
$$
\hspace {-44mm}d\zeta = \displaystyle\frac{xdx}{\sqrt{\varepsilon(x)}} =
\displaystyle\frac{dy}{2\sqrt{\varepsilon(y)}}, 
\eqno{(10)}
$$
Eq.(9) is rewritten in a self-conjugate form and becomes the equation for a nonlinear oscillator

$$
\hspace {-46mm}G^{\prime \prime}_{\zeta \zeta} +{\Omega^2 (\zeta (x))} G = 0, 
\eqno{(11)}
$$
where $\;y = x^2;$ $\;\Omega^2\;$ is the squared "frequency". 

Eq.(10) not always can be integrated in elementary functions, and more simply the function $\;\Omega^2\;$ is written as a function of the variable $y$:

$$
\hspace {-36mm}\Omega^2 = - d(\Phi/y)dy - 2\chi W_{el}/y. 
\eqno{(12)}
$$

Taking into account the contribution of the electric field energy, the function $\Phi$ playing the role of the Newtonian gravitational potential inside the star can be obtained from the gravitational equations as 

$$
\hspace {-28mm}\Phi = 1 - \varepsilon = \displaystyle\frac{\chi}{x} \int (\mu(x)+W_{el}) x^2 dx 
$$
$$
= \displaystyle\frac{\chi}{2\sqrt{y}} \int (\mu(y)+W_{el})\sqrt{y}dy. 
\eqno{(13)}
$$

Choosing the mass density $\;\mu\;$ and the electric field energy density inside the star 
$\;W_{el}\;$ as square-law functions of $\;x\;$ (as was made in [3]),
$$
\hspace {-38mm}\mu =  \mu_0 (1-x^2) = \mu_0 (1-y); 
\eqno{(14)}
$$

$$
\hspace {-46mm}W_{el} = \displaystyle\frac{\lambda^2 R^2 x^2}{8\pi} =
\displaystyle\frac{\Lambda^2 y}{8\pi}, 
\eqno{(15)}
$$
we obtain from (11) the equation for a linear oscillator

$$
\hspace {-56mm}G^{\prime \prime}_{\zeta \zeta} +  \Omega_0^2  G = 0 
\eqno{(16)}
$$
with a constant value of the natural frequency 

$$
\hspace {-48mm}\Omega_0^2 =
\displaystyle\frac{\chi}{5} (\mu_0 -\displaystyle\frac{11\Lambda^2}{8\pi}),
\eqno{(17)}
$$

The continuity of the electric energy density on the stellar surface requires 
 $\Lambda = Q/R^2,\;$ where $\;Q\;$ is the integral electric charge of the star and $\mu_0$ is its central mass density.

In this case the function $\Phi$ may be written as 
$$
\hspace {-34mm}\Phi =\chi y (\displaystyle\frac{\mu_0}{3}+
\displaystyle\frac{1}{5}(\displaystyle\frac{\Lambda^2}{8\pi} -\mu_0) y).
\eqno{(18)}
$$

The radial dependence of the electric charge density in the interior field of the star can be found for the chosen functional law of the observable electrical energy density from Maxwell's equation (8.4):

$$
\hspace {-48mm}\rho(x) = Q \cdot \sqrt{\varepsilon(x)}/V, 
\eqno{(19)}
$$
where $V = 4\pi R^3/3$ is the Euclidean volume of 3-space. 

The general solution of Eq.(16) for the required function $G$ will be written as 

$$
\hspace {-48mm}G = G_0 \cos(\Omega_0 \zeta + \alpha), 
\eqno{(20)}
$$
and for the metric function $g_{00} = F = G^2$ 

$$
\hspace {-42mm}F = G_0^2 \cos^2(\Omega_0 \zeta(x) + \alpha), 
\eqno{(21)}
$$
where $\zeta(x)$ is a quadrature depending on the relation between the parameters $\mu_0$ and $\Lambda.$

\end{document}